\documentclass[11pt]{article}
\usepackage[pctex32]{graphics}
\usepackage{amssymb}
\usepackage{latexsym}
\usepackage{epsfig} 
\usepackage{multirow} 
\usepackage{pstcol}
\usepackage{float}
\usepackage{epstopdf}
\usepackage{mathrsfs}

\usepackage{amsmath}
\usepackage{amssymb}
\usepackage{relsize}
\usepackage{multirow}

\usepackage{graphicx}
\usepackage{multicol} 

\usepackage{times}
\usepackage{epstopdf}

\definecolor{Red}{rgb}{1,0.,0.}
\usepackage{fancyhdr}
\usepackage{setspace}
\newcommand{\R}{{\mathbb R}}

\newcommand{\mC}{{\mathsf C}}
\newcommand{\mI}{{\mathsf I}}

\title{Bayesian dynamical estimation of the parameters of an SE(A)IR COVID-19 spread model}
\author{D Calvetti$^{1,2}$ \and A Hoover$^3$ \and J Rose$^2$ \and E Somersalo$^{1,2}$}
\date{$^1$ Department of Mathematics, Applied Mathematics, and Statistics\\
Case Western Reserve University\\
$^2$ Center for Community Health Integration\\
Case Western Reserve University\\
$^3$ Department of Mathematics \\
University of Akron}

\begin{document}
\maketitle

\begin{abstract}
In this article, we consider a dynamic epidemiology model for the spread of the COVID-19 infection. Starting from the classical SEIR model, the model is modified so as to better describe characteristic features of the underlying pathogen and its infectious modes. In line with the large number of secondary infections not related to contact with documented infectious individuals, the model includes a cohort of asymptomatic or oligosymptomatic infectious individuals, not accounted for in the data of new daily counts of infections. A Bayesian particle filtering algorithm is used to update dynamically the relevant cohort  and simultaneously  estimate the transmission rate
as new daily incidence and mortality data become available. The underlying assumption of the model is that the infectivity rate is dynamically changing, either because of a mutation of the pathogen or in response to mitigation and containment measures. The sequential Bayesian framework naturally provides a quantification of 
of uncertainty in model parameter estimates, including the reproduction number, and of the size of the different cohorts. Moreover, we introduce  a dimensionless quantity, which is the equilibrium ratio between asymptomatic and symptomatic cohort sizes, and propose a simple formula to estimate the quantity. This ratio leads naturally to another dimensionless quantity that plays the role of the basic reproduction number $R_0$ of the model. When we apply the model and particle filter algorithm to COVID-19 infection data from several counties in Northeastern Ohio and Southeastern Michigan we found the proposed reproduction number $R_0$ to have a consistent dynamic  behavior within both states, thus proving to be a reliable summary of the success of the mitigation measures. 
\end{abstract}

\section{Introduction}

Since its emergence in Wuhan, China, at the end of 2019,  the novel coronavirus SARS-CoV-2 has spread worldwide. In a little over four months it has evolved into a pandemic affecting nearly every country in the world in spite of the measures taken to control and contain the contagion. The novelty of the virus, which is related, but different from the coronaviruses responsible for the SARS  and MERS epidemics in 2003 and 2009, poses a challenge to using mathematical models to predict the dynamics of the pandemic and the effects of changing mitigation strategies.  Following the initial outbreak, number of contributors have either proposed mathematical models specifically designed to examine the spread of COVID-19 \cite{kucharski2020early, leung2020first},  or addressed important points related to the estimation of the key model parameters \cite{anastassopoulou2020data, zhao2020preliminary} from partial, biased daily updated data \cite{johns2020coronavirus, USAfacts} and incomplete information about the pathogen responsible for the pandemic. 

Early virological assessment of SARS-CoV-2 \cite{Wolfel2020} from a small sample of patients in Germany found active virus replication in the upper respiratory tract, as opposed to SARS where the predominant expression was in the lower respiratory tract. Patients were tested when symptoms were mild or in the prodromal phase. Pharyngeal viral shedding was very high during the first week of symptoms, peaking on the fourth day, with shedding continued after the symptoms subsided, well into the second week. Moreover, most of the patients seemed to have passed their shedding peak in the upper respiratory tract at the time of the first  testing, suggesting efficient transmission of SARS-CoV2 through pharyngeal shedding when the symptoms are mild. In \cite{He2020} peak shedding is suggested to occur around day 0, and $44\%$ of the transmission is estimated to occur in the asymptomatic phase. These specific clinical features need an interpretation in the epidemic models.

The spread and speed of the COVID-19 pandemic has triggered a burst of modeling activity in the effort to help predict the location and intensity of the next hotspots. One of the most used indicators of the potential for spread of an infectious agent is the basic reproduction number $R_0$ \cite{macdonald1952analysis, macdonald1957epidemiology, heesterbeek2002brief}, although questions have been raised as to how it should be computed \cite{Diekmann,diekmann2010construction, dietz1993estimation} and interpreted \cite{Delamater, Heffernan, li2011failure}, and what can be inferred from it \cite{Ridenhour}.  In general, $R_0>1$ signals epidemic potential, and the larger the $R_0$, the faster the spread of the infectious agent. In \cite{Wang}, where a SEIR model is calibrated and tested on the Wuhan outbreak data, the initial estimate of  $R_0$, whose definition and significance will be discussed later, was updated after the implementation of strict measures for control and prevention of the contagion.  The model was then used to predict the magnitude of the pandemic by predicting the number of infected individuals on February 29, 2020 under the two different scenarios of an increasing $R_0$, changing from 1.9 to 2.6 and ending at 3.1, corresponding to letting the pandemic follow its course, and that of a decreasing $R_0$, going from 3.1 to 2.6, then 1.9, and eventually 0.9 and 0.5, following the mitigation measures taken in Wuhan.

Of models particularly adapted to descibe COVID-19, we mention the SIDARTHE model \cite{giordano2020modelling}, consisting of eight separate compartments to address the different role in the spread of the pandemic of unreported infectious individuals, who presumably constitute the great majority. The model parameters, inferred from official data on the number of diagnosed and recovered cases, and fitted by recursive least squares methods, are adaptively changed during the simulation to reflect the introduction of progressively restrictive measures. More specifically, for the Wuhan outbreak the value of $R_0$, set to 2.38 on day 1, is changed to 1.6 on day 5, increasing to 1.8 on day 12 and returning to 1.6 on day 22. After day 22, $R_0$ is reduced to 0.99, and from day 38 it is set to 0.85. The SIDARTHE model is then used to predict the course of the pandemic if the lockdown measures are either weakened or tightened.

The importance of considering the contribution from undocumented infections is repeatedly addressed in the COVID-19 literature \cite{LancetWu}. The population seroprevalence of antibodies in a random sample of 3300 people in Santa Clara County, California  \cite{bendavid2020covid} early in the U.S. outbreak, would indicate the number of infections to be much higher than the number of reported cases, possibly by a factor of  50 to 80. This is in line with 10\% seropositivity reported in the town of Robbio, Italy, and 14\% in the town of Gangelt,  Germany. 

Another important issue, raised in \cite{tsang2020effect}, is  the bias in the estimate of key epidemic parameters if the delay in case reporting is not taken into consideration.

Documented and undocumented infections are accounted for in separate compartments in the SEIR network dynamic metapopulation model in \cite{Science_Li},  which assumes that asymptomatic, or oligosymptomatic cases can expose a far larger proportion of the susceptible cohort to the virus than can reported cases, with asymptomatic cases less infectious than symptomatic ones. The model takes into account the travels of infectious individuals between cities by setting up a network comprised of 375 cities in China, with the amount of traffic between any pair of cities estimated from mobility data, and assigns fours separate cohort to each city. The estimator of the distribution of the six model parameters via a variant of ensemble Kalman filter suggests that in the first month of the Wuhan outbreak, only approximately  one out of six infections was reported. The adoption of strong mitigation measures starting from January 23, 2020 is included in the model by first reducing the mobility coefficients to 2\%, of the normal value, and by setting them to zero when all traffic stopped. The median latent period of COVID-19 in Wuhan according to this model is 3.7 days, and the median infectious period 3.48 days, with a median delay in case reporting of 10 days at first, and 6 days after the lockdown. The median estimate of the reproduction number varies from 2.38 in the early phase, to 1.66 and 0.99 after containment measures, when the fraction of undocumented infections decreases sharply.

In \cite{lourencco2020fundamental}, the epidemic is assumed to go through three phases: in the first there is a slow accumulation of new infections, most of which are unreported. The second phase is characterized by a rapid growth of infections, diseases and deaths, while in the third phase the epidemic slows down due to the depletion of susceptible individuals. The SIR model used to describe the course of the COVID-19 pandemic, calibrated on the daily number of deaths in the United Kingdom and Italy and with $R_0=2.75$ and $R_0=2.225$, respectively, inferred from the literature, suggest that the epidemic already started one month before the first reported death.

The most widely used tools in the United States for predicting the dynamics of the pandemic include the CHIME \cite{abdulrahman2020simcovid,weissman2020locally} and the IHME forecasting model \cite{Murray2020}. The former is an SIR/SEIR based differential-algebraic simulation tool with an accessible interface.
The latter, in order to counteract the typical SIR overestimation of the proportion of infected and to focus on the most severely ill patients, uses empirically observed COVID-19 
population mortality curves. The underlying  statistical model assumes that the cumulative death rate at each location follows the Gaussian error function, and produces long and short term predictions.

The goal of this article is to provide a relatively simple and flexible model equipped with a Bayesian state and parameter estimation protocol to dynamically process the COVID-19 data inflow to assess the current standing of the population and to make short term forecasts of the progression. All parameters and outputs of the algorithm are easily interpretable and adjustable. The underlying model is an adaptation  of the classical SEIR model \cite{Kermack}, adjusted to better conform with certain specific features of the current COVID-19 epidemics. In particular, we re-interpret the cohort of exposed individuals, defining it as individuals who carry and shed the virus asymptomatically, presymptomatically, or oligosymptomatically, thus not being isolated or  hospitalized. Moreover, this cohort does not contribute to the number of confirmed cases. We refer to this model as SE(A)IR.
We propose a Bayesian particle filtering (PF) algorithm for estimating dynamically the state vector consisting of the sizes of the four cohorts in the model based on a Poisson distributed observation of the infected cohort size, with the dynamic model generating the mean of the distribution.  Concomitantly, we estimate the presumably dynamically changing rate of transmission with posterior envelopes of model uncertainty.  Being a fully Bayesian algorithm, the output consist of model uncertainties. Moreover, we show the forecasting of the expected number of new infections based on the model with predictive uncertainty envelopes.

One key factor contributing to the challenge of making predictions and planning is the unknown number of individuals spreading the virus asymptomatically. The particle filter algorithm provides an estimate for the ratio of the asymptomatic and symptomatic virus carriers. A novel feature of this contribution is the derivation of a Riccati type equation for the ratio of the sizes of the two cohorts. Moreover, the Riccati equation has a short time approximate stable equilibrium. The equilibrium value, which can  be analytically calculated from the model parameters, corresponds well to the model-based estimated ratio and  can be used to define a dynamically changing effective basic reproduction number $R_0$ for the epidemic, facilitating the comparison of model predictions with other models.

The methodology is extensively tested using COVID-19 data of 18 counties in Northeastern Ohio (Cleveland area) and 19 counties in Southeastern Michigan (Detroit area) during the period from early March 2020 to early May 2020, including the period when both states introduced similar, yet slightly different mitigation protocols.

\section{COVID-19 epidemiology model}

Compartment models in mathematical epidemiology partition a homogenous and well-mixed population into cohorts of individuals at different stages of the infection \cite{calvetti2012computational}. The  popular SIR model, with separate compartments for susceptible (S), infected (I) and recovered (R),  introduced nearly a century ago by Kermack and McKendrick \cite{Kermack}, introduced a population dynamics component into the previous, purely phenomenological statistical models (see, e.g. \cite{Farr}) that still seem to have a life of their own in modeling the COVID-19 epidemic \cite{Murray2020}. 

A significant challenge for the control and containment of the COVID-19 epidemic is the spread of the infection by a large portion of asymptomatic or lightly symptomatic infectious individuals who are unaware of being vectors of the virus.  This is especially problematic when, as is the case at the time of the writing of this article, due to limited availability, testing  priority is given to symptomatic individuals or to vulnerable populations, thus the size of the asymptomatic cohort must be estimated indirectly. In the next subsection, we propose a compartment model that can be used to obtain such an estimate, and discuss its advantages and limitations.

\subsection{SE(A)IR model}

\begin{figure}
\centerline{
\includegraphics[height=4cm]{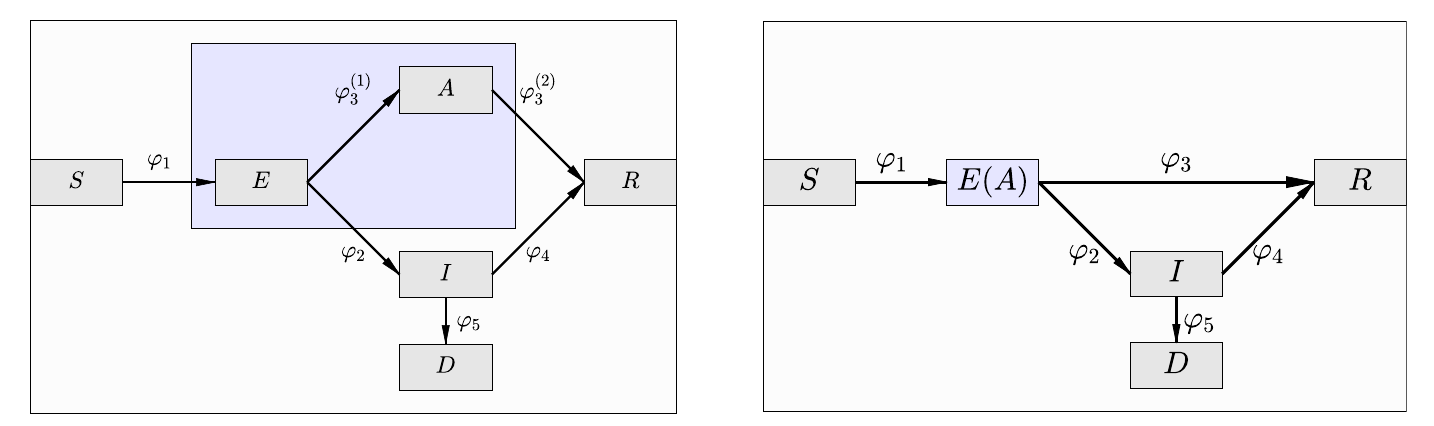}
}
\caption{\label{fig:compartment} Left: compartment diagram of the model including both symptomatic and asymptomatic infected cohorts. After a non-infectious incubation period, the exposed individuals branch either to symptotmatic ($I$) or asymptomatic ($A$) infectious compartments, with unknown frequencies and unknown reasons. 
In the modified SE(A)IR model (right), the two compartments $E$ and $A$ on the blue background are lumped together to form the fictitious $E(A)$ compartment representing the exposed, asymptomatic cohorts. The two fluxes $\varphi_3^{(1)}$ and $\varphi_3^{(2)}$ are merged to a single flux $\varphi_3$. }
\end{figure}

We begin by considering a modification of the classical SEIR model where the infected cohort $I$ is subdivided into two groups according to the manifestation of symptoms, denoting by $A$ the asymptomatic, infected, and infectious  subcohort and by $I$ the symptomatic infected one.  Hence, while $E$ and $A$ are both asymptomatic and technically infected, the $E$ cohort is not infectious as in the classical SEIR model, as opposed to $A$ that sheds the virus. 
The compartment model, schematically represented by the branching flow diagram in the left panel of Figure~\ref{fig:compartment}, is governed by the system of differential equations 
\begin{eqnarray}\label{big system}
\frac{dS}{dt} &=& -\varphi_1, \nonumber \\
\frac{dE}{dt} &=& \varphi_1 -\varphi_2 - \varphi_3^{(1)}, \nonumber \\
\frac{dA}{dt} &=& \varphi_3^{(1)} - \varphi_3^{(2)}, \\
\frac{dI}{dt} &=&  \varphi_2 -\varphi_4-\varphi_5,\nonumber \\
\frac{dR}{dt} &=& \varphi_3^{(2)} +\varphi_4, \nonumber
\end{eqnarray}
where the functional form of the fluxes is will be specified below. The COVID-19 data, consisting of the daily count of newly reported infection, corresponds to observations of the flux $\varphi_2$: In the absence of additional population-level inputs, this is the data that must be used to estimate the cohort sizes and model parameters.
In particular, if no data concerning the asymptomatic cohort dynamics are available, the values of the fluxes $\varphi_3^{(1)}$ and $\varphi_3^{(2)}$ can be set rather arbitrarily, since they are minimally connected with the fluxes in the lower branch of the flow diagram, whose values are part of the observations. To estimate for the size of the asymptomatic cohort in the absence of additional information, it is necessary to modify the model. Below we propose a modified version of the model that is suitable for estimating the size of the asymptomatic cohort, while retaining many of the salient features of the extended model. The modification that we propose applies an approach similar to that used in metabolic network model reduction \cite{marangoni2003enzyme}, where lumping enzymatic reactions whose parameters cannot be estimated from the data is fairly common.

In our reduced model, the fictitious compartment $E(A)$ embeds the asymptomatic cohort into the exposed one, as depicted in Figure\ref{fig:compartment}, (right), thus the size of the new cohort is
\[
 E(A) = E + A,
\]
and, from (\ref{big system}) it follows that
\[
 \frac{d E(A)}{dt} = \frac{dE}{dt}+\frac{dA}{dt} = \varphi_1 -\varphi_2 - \varphi_3^{(2)}.
\]
In this manner, the flux $\varphi_3^{(1)}$, that describes a variation internal to the lumped compartment, is no longer part of the model governing equations. Assuming, for simplicity, that both asymptomatic and symptomatic individuals move to the recovered compartment with the same relative rate, denoted by $\gamma$, it follows that
\[
 \varphi_3  = \gamma E(A), \quad \varphi_4 = \gamma I.
\] 
The flux $\varphi_2$ represents the rate at which part of the exposed individuals develop symptoms, and for simplicity, we assume that the flux is proportional to the size of the compounded cohort, that is,
\[
\varphi_2 = \eta E(A).
\]
One of the most relevant parameters in an epidemic is the rate at which susceptible individual are infected. In the classical SEIR model, the transmission flux assumes the form
\[
\varphi_1 = \beta \frac{I}{N} S,
\]
where $\beta$ is the transmission rate and $N$ is the population size. However, one of the problems of  COVID-19 is the significant amount of virus transmission by the asymptomatic cohort within $E(A)$.  Our model assumes that in the current COVID-19 epidemics, most infected individuals who have developed symptoms, are either hospitalized or in self-isolation, thus playing a limited role in the exposure of susceptible individuals to the contagion.  Acknowledging the differing roles that symptomatic and asymptomatic virus shedders play in transmission to susceptibles, we define
\[
 \varphi_1 = \beta \frac{p E(A) + q I}{N} S,
\]
where $p$ and $q$ are frequencies,  $0\leq p, q\leq 1$. The frequency $p$ may be interpreted as the fraction of the compound cohort E(A) who are infectious. Classical SEIR models, where usually it is assumed that the exposed cohort is not infective, hence only the infected $I$ cohort is responsible for the spreading of the epidemic, can be accounted for by setting $p=0$, $q=1$. 
In our model  COVID-19 dynamics, we assume that $p>q$.
%and in the simulations use  the values $(p,q) = (1,0.1)$. 
The transmission rate $\beta$, which is the quantity of primary interest when it comes to assessing how fast the epidemics spread, integrates elements related to the characteristics of the pathogen determining the probability of infection of a given susceptible contact, as well as factors related to social behavior, including the number and nature of daily contacts. 
In summary, the goverining equations of the proposed COVID-19 model are
\begin{eqnarray}
\frac{dS}{dt} &=& -\beta \frac{p E(A) + q I}{N} S ,  \label{S}\\
 \frac{dE(A)}{dt} &=& \beta \frac{p E(A) + q I}{N} S - \eta E(A)  - \gamma E(A), \label{E}\\
 \frac{dI}{dt} &=&  \eta E(A)  - \gamma I - \mu I,\label{I}\\ 
  \frac{dR}{dt} &=&   \gamma E(A) + \gamma I, \label{R}
\end{eqnarray}  
where $\beta$ is the transmission rate, $\gamma$ is the recovery rate, $\eta$ is the symptomatic rate integrating in it the incubation process $E\rightarrow I$, and $\mu$ is the death rate.

The data that is used to inform our parametrized model, as well as to update the state estimation of the cohorts over time is the number of confirmed, symptomatic new daily cases, or, equivalently, daily observations of the flux
\[
 \varphi_2 =\eta E(A),
\]
or, in fact, of a stochastic integer-valued realization of it, as explained in more detail in the next section.

Since, as already pointed out, the rate of transmissibility integrates not only properties of the pathogen, but also factors related to human behavior that change in the course of an epidemic, in general $\beta$ is not a static  parameter.  In the filtering approach used to estimate the size of the cohorts in the course of the epidemic, to be explained below, $\beta$ is modeled as a time dependent parameter. In the absence of a known deterministic dynamical evolution model for $\beta$, its proposed changes will be described in terms of a stochastic geometric random walk.

Before describing the computational methodology on which our model predictions are based, some qualitative comments about the model are in order. In classical SEIR models, the inclusion of the exposed cohort $E$ adds a time delay corresponding to the incubation period not accounted for in SIR models. While our expression for $\varphi_1$ implicitly assumes that the infected and asymptomatic individuals are immediately infectious, the flux $\varphi_2$ introduces a slight delay in the $I$ cohort if, for instance, the transmission rate is changed. Moreover, if $p\neq 0$, we may write
\[
 \varphi_1 = \beta p\frac{E(A) + (q/p) I}{N} S,
\]
and by absorbing $p$ into the transmission rate $\beta$, the model can be written in terms of the ratio $q/p$ determining the size of the relative effect of infected symptomatic individuals on the spread of the infection. 
Hence, without a loss of generality, we may set $p=1$, and assume $q<1$. The value $q = 0.1$ used in our numerical tests is a rough estimate, and the effect of that parameter on the estimated quantities is briefly discussed in the results. Finally, we point out that the simplifying assumption that the symptomatic and asymptomatic individuals recover at the same relative rate is not essential for the methodology developed below, and can be easily removed.

In the discussion to follow, we simplify the notation and write $E$ instead of $E(A)$ for the exposed/asymptomatic cohort size.   

\section{Bayesian particle filtering}\label{sec:PF}

Bayesian filtering algorithms update the information about the state vector and parameters in a sequential manner as new data arrives. In the following, we use the convention of denoting by upper case letters the random variables and by lower case letter their realizations. We refer to \cite{liu2001combined,kaipio2006statistical} for particle filtering, and to \cite{arnold2014parameter,arnold2013linear,arnold2015astrocytic} for the particular type of applications to ODE systems.

Let $\{X_t\}$ denote a discrete time Markov process, with the transition probability distribution $\pi_{X_{t+1}\mid X_t}(x_{t+1}\mid x_{t})$. Furthermore, let $\{B_t\}$ denote the stochastic process representing the observations, and let $\pi_{B_t\mid}(b_t\mid x_t)$ denote the likelihood density, where we implicitly assume that the current observation $B_t$ depends only on the current state $X_t$ and not on the past. Finally, let ${\mathscr B}_t$ denote the cumulative data up to time $t$, that is,
\[
 {\mathscr B}_t = \{B_1,B_2,\ldots,B_t\}.
\]
We denote by ${\mathcal B}_t$ the set of observed realizations, ${\mathcal B}_t = \{b_1,b_2,\ldots,b_t\}$.
In Bayesian filtering algorithms the update of the posterior distributions is carried out in two consecutive steps, 
\[
 \pi_{X_t\mid {\mathscr B}_t} \rightarrow  \pi_{X_{t+1}\mid {\mathscr B}_t} \rightarrow  \pi_{X_{t+1}\mid {\mathscr B}_{t+1}},
\] 
where the first step is referred to as propagation step, and the second as correction, or analysis step. The first step is accomplished through the Chapman-Kolmogorov formula,
\[
 \pi_{X_{t+1}\mid {\mathscr B_t}}(x_{t+1}\mid  {\mathcal B_t}) = \int  \pi_{X_{t+1}\mid X_t}(x_{t+1}\mid x_{t})  \pi_{X_{t}\mid {\mathscr B_t}}(x_{t}\mid  {\mathcal B_t})  dx_t,
\]
while the analysis step builds on Bayes' formula,
\[
  \pi_{X_{t+1}\mid {\mathscr B_{t+1}}}(x_{t+1}\mid  {\mathcal B_{t+1}}) =  \pi_{X_{t+1}\mid {\mathscr B_t}}(x_{t+1}\mid  {\mathcal B}_t) \pi_{B_{t+1}\mid X_{t+1}}(b_{t+1}\mid x_{t+1}),
\]
that is, the predicted distribution of $X_{t+1}$ acts as the prior as the next observation arrives. In the following, we specify the transition probability kernel and the likelihood in our model, and describe the computational steps for the numerical implementation. We then extend the discussion to include the estimation of static parameters.

Consider our modified SE(A)IR model introduced in the previous section, and define the state vector at time $t$ 
\[
 z_t = \left[\begin{array}{c} S_t \\ E_t \\ I_t \\ R_t\end{array}\right], \quad t=1,2,\ldots.
\]
Since we assume that the infectivity parameter $\beta$ may vary over time, we denote its value at time $t$ by $\beta_t$. 
Formally, let  $\psi$ be the numerical propagator advancing the state variable and the infectivity parameter from one day to the next, $t\to t+1$,
\[
 \psi(z_t,\beta_t) = z_{t+1}.
\]
In our computations, the time integration performed by means of a standard ODE solver such as Runge-Kutta, and during the one day propagation step, the infectivity $\beta_t$ is kept constant. 

Formally, we write a propagation model of the form
\begin{equation}\label{propagate}
 x_t = \left[\begin{array}{c}  z_t \\ \beta_t\end{array}\right] \rightarrow F(x_{t}) = \left[\begin{array}{c}  \psi(z_t,\beta_t) \\ \beta_t\end{array}\right] = \widehat x_{t+1},
\end{equation}
and we account for uncertainties both in the state vector $z_t$ and the parameter vector $\beta_t$ by introducing an innovation term. We guarantee that all components of the state vector and the parameter $\beta$ are nonnegative, with a multiplicative innovation, assuming a geometric random walk model,
\begin{equation}\label{innovation}
 \log X_{t+1} \sim {\mathcal N}(\log\widehat x_{t+1}, \mC),
\end{equation}
where $\mC\in\R^4$ is a diagonal positive definite matrix.  

In the definition of the likelihood, we assume that  the data consist of realizations $b_t$ of the daily new infections, denoted by $B_t$. The new infection count is assumed to be Poisson distributed, with the Poisson parameter equal to the flux $\varphi_2(t)$, 
 \[
   B_t \sim{\rm Poisson}(\varphi_2(t)), \quad \mbox{where} \quad  \varphi_2(t) = \eta E_t.
\]
Therefore, the likelihood of the observed number $b_t$ of new infections is 
\[
 \pi_{B_t\mid Z_t}(b_t\mid z_t) = \frac{(\eta E_t)^{b_t}}{b_t!} e^{-\eta E_t}.
\] 
To initialize the process, we need to define the initial state at the time $t=0$ before the first observed infections. Since the initial state is unknown, its initial value becomes part of the estimation problem. We postulate that before the first infection is observed, there is an unknown number of asymptomatic individuals in the community. We assume that the initial number of asymptomatic cases follows a Poisson distribution with a uniformly distributed expected value $\Lambda \sim{\rm Uniform}([0,\lambda_{\rm max}])$. Moreover, we  assume that $\beta_0$ follows a uniform distribution over some interval $[\beta_{\rm min},\beta_{\rm max}]$.

We are now ready to outline the particle filter (PF) algorithm for estimating the state vector $X_t$ based on the infection count. Assume that at time $t$ a sample from the distribution $\pi_{X_t\mid{\mathscr B}_t}$ 
\[
 \{x_t^1,x_t^2,\ldots,x_t^N\},
\]
is available, and each sample point is associated its corresponding weight $w_t^j$.  We write a particle approximation of the Chapman-Kolmogorov formula 
\begin{eqnarray*}
 \pi_{X_{t+1}\mid {\mathscr B_t}}(x_{t+1}\mid  {\mathcal B_t}) &=& \int  \pi_{X_{t+1}\mid X_t}(x_{t+1}\mid x_{t})  \pi_{X_{t}\mid {\mathscr B_t}}(x_{t}\mid  {\mathcal B_t})  dx_t  \\
 & \approx & \sum_{j=1}^N w_t^j \pi_{X_{t+1}\mid X_t}(x_{t+1}\mid x_{t}^j), 
\end{eqnarray*}
and combine it with Bayes' formula to obtain the updating formula
\[
 \pi_{X_{t+1}\mid {\mathscr B_{t+1}}}(x_{t+1}\mid  {\mathcal B_{t+1}}) \approx \sum_{j=1}^N w_t^j \pi_{X_{t+1}\mid X_t}(x_{t+1}\mid x_{t}^j) \pi_{B_{t+1}\mid X_{t+1}}(b_{t+1}\mid x_{t+1}).
\]
Let $\widehat x_{t+1}^j$ denote a propagated predictor particle associated with $x_t^j$, that is,
\[
 \widehat x_{t+1}^j = F(x_t^j).
\]
At the arrival of the next observation $b_{t+1}$, the likelihood  $\pi_{{\mathscr B}_{t+1}\mid X_{t+1}}(b_{t+1}\mid \widehat x_{t+1}^j) $ expresses how well the predictor is at  explaining the data. Writing the above formula as
\[
 \pi_{X_{t+1}\mid {\mathscr B_{t+1}}}(x_{t+1}\mid  {\mathcal B_{t+1}}) \approx \sum_{j=1}^N \underbrace{\left\{w_t^j \pi_{B_{t+1}\mid X_{t+1}}(b_{t+1}\mid \widehat x_{t+1})\right\}}_{(1)}
\underbrace{\left\{ \frac{\pi_{B_{t+1}\mid X_{t+1}}(b_{t+1}\mid x_{t+1})}{\pi_{B_{t+1}\mid X_{t+1}}(b_{t+1}\mid \widehat x_{t+1})}\right\}}_{(2)}
 \underbrace{\pi_{X_{t+1}\mid X_t}(x_{t+1}\mid x_{t}^j)}_{(3)},
\] 
we can interpret the formula as follows: Given a predictor particle $\widehat x_{t+1}^j$, the expression (1), combining the importance of its predecessor in the weight and its fitness in the likelihood, evaluates the relevance of the particle;  (2) weighs the importance of the new particle relative to the predictor, and the transition kernel (3)  generates a new particle from the predictor by the formula (\ref{innovation}). This hierarchical organization is the backbone of the particle filter algorithm.
 
\bigskip
\hrule
\medskip

{\bf Particle filtering algorithm}
\bigskip

\hrule

\begin{itemize}
\item[] {\bf Initialize:} Draw $N$ independent  realizations of $\beta_0\sim{\rm Uniform}([\beta_{\rm min},\beta_{\rm max}])$ and $\Lambda  \sim{\rm Uniform}([0,\lambda_{\rm max}])$,
\[
 \{\beta_0^1,\beta_0^2,\ldots,\beta_0^N\}, \quad  \{\lambda^1,\lambda^2,\ldots,\lambda^N\},
\]
and generate $N$ realizations of the initial state of $Z_0$,
\[
 z_0^j = \left[\begin{array}{c} N - \lambda^j \\ \lambda^j \\ 0 \\ 0\end{array}\right], \quad j = 1,2,\ldots,N,
\]
then define the initial cloud of the particles
\[
\{x_0^1,x_0^2,\ldots,x_0^N\}, \quad  x_0^j = \left[\begin{array}{c} z_0^j \\ \beta_0^j\end{array}\right].
\]
Set $t=0$.   
\item[] {\bf While $t < t_{\rm max}$ do}
\begin{itemize}
\item[(a)] Propagate the particles according to  (\ref{propagate}) to generate the predictive particle cloud
\[
\{ \widehat x_{t+1}^1,\widehat x_{t+1}^2, \ldots, \widehat  x_{t+1}^N\}.
\]
\item[(b)] Extract the second component $\widehat e_{t+1}^j$ of each of the propagated particles, and compute the weights,
\[
  \widehat g^j_{t+1} = w_t^j \frac{(\eta \widehat E_{t+1}^j)^{b_{t+1}}}{b_{t+1}!} e^{-\eta \widehat E_{t+1}^j}, \quad  \widehat g^j_{t+1} \leftarrow \frac{\widehat g^j_{t+1}}{\sum_\ell \widehat g^\ell_{t+1}}.
\]  
\item[(c)] Sample with replacement $N$ indices $\ell_j\in\{1,2,\ldots, N\}$, $j=1,2,\ldots,N$, with the probability weights $g^j_{t+1}$. Define the new resampled predictive cloud 
\[
 \widehat x_{t+1}^j \leftarrow \widehat x_{t+1}^{\ell_j}.
\]
Generate a new particle cloud through the innovation process,
\[
 \log x_{t+1}^j = \log\widehat x_{t+1}^j + \mC^{1/2} w^j, \quad w^j\sim{\mathcal N}(0,\mI_5).
\]  
\item[(d)] Extract the second component $e_{t+1}^j$ from each new particle, and compute the new likelihood weights,
\[
 g^j_{t+1} = \frac{(\eta E_{t+1}^j)^{b_{t+1}}}{b_{t+1}!} e^{-\eta  E_{t+1}^j}, \quad   g^j_{t+1} \leftarrow \frac{ g^j_{t+1}}{\sum_\ell  g^\ell_{t+1}}.
\] 
Update the weights,
\[
 w_{t+1}^j =  \frac{g_{t+1}^j}{\widehat g_{t+1}^j}.
\] 
Advance $t\to t+1$.
\end{itemize}
\item[] {\bf end do}
\end{itemize}
\hrule

In the particle filter algorithm, the model parameters $\gamma$, $\eta$ and $\mu$ are fixed. It is possible to modify the algorithm to estimate these parameters also. In that case, to estimate the death rate $\mu$, the information about the deceased must be included in the data, as the new infection count is essentially insensitive to that parameter. The parameters $\gamma$ and $\eta$ are less time-dependent than the infectivity $\beta$, and we set their values according to what is suggested in the literature.

\subsection{Ratio of asymptomatic to symptomatic cohorts}

A challenges in forecasting the spread of the COVID-19 epidemic is the presumably large portion of population that is asymptomatic or shows only light symptoms, while shedding the virus, which in our model is part of the cohort $E$. The ratio between the numbers of symptomatic and asymptomatic individuals,
\[
 \rho(t) = \frac{E(t)}{I(t)},
\]
can be estimates from the output of the particle filter. Differentiating $\rho$ with respect to time,
\[
 \frac{d\rho}{dt} = \frac 1I\frac{dE}{dt} - \frac{E}{I^2} \frac{dI}{dt},
\]
expressing the derivatives of $E$ and $I$ in terms of the governing equations (\ref{E})-(\ref{I}), and simplifying, we find that $\rho$ satisfies the Riccati equation
\begin{eqnarray*}
   \frac{d\rho}{dt}  &=& \beta\frac{p+q\rho}{N} S - (\eta+\gamma)\rho - \eta \rho^2 -(\gamma+\mu)\rho \\ 
   &=& -\eta\left(\rho^2 + \left(\frac{\eta+2\gamma+\mu}{\eta} - \frac\beta\eta\frac SN q\right)\rho - \frac\beta\eta\frac SN p\right).
\end{eqnarray*}
If we assume that the frequency of susceptible cohort $\nu_S = S/N$ is approximately constant, we find an equilibrium state $\rho^*$ of the ratio,
\begin{equation}\label{equilibrium}
 \rho^* = -\frac 12   \left(\frac{\eta+2\gamma+\mu}{\eta} - \frac\beta\eta\nu_S q\right) + \sqrt{\frac 14  \left(\frac{\eta+2\gamma+\mu}{\eta} - \frac\beta\eta\nu_S q\right)^2
 +\frac\beta\eta\nu_S p},
\end{equation}
which is a stable equilibrium.
In particular, when the infection has not spread yet to  a significant portion of the population, substituting  $\nu_S\approx 1$ yields a potentially useful estimate for the prevalence of the asymptomatic infection in the population.

In the computed results, a comparison of the values of ratio $\rho$ computed from the state vectors and the equilibrium estimate, can be used to asses whether the infection pattern is settling to or near the equilibrium state.

The equilibrium condition provides a natural way of defining a basic reproduction number $R_0$ for the model. Assuming equilibrium, we have
\[
 \frac{d I}{dt} = \eta E - (\gamma+\mu)I = -(\gamma + \mu)\left(1 - \frac{\eta}{\gamma + \mu}\rho^*\right) I = -(\gamma + \mu)(1-R_0) I,
\]
where
\[
 R_0 =  \frac{\eta}{\gamma+\mu} \rho^*.
\] 
Thus,  when $R_0=1$ the {\em infection stops growing} under the assumption of a stable ratio of asymptomatic to symptomatic infections. This interpretation of the reproduction number is in close agreement with the $R_0$ for SIR models. While not perfect, this number summarizes the phase of the epidemic,  as our computed examples will demonstrate.

\section{Results}

To test our model, we consider the COVID-19 data of new infection counts in different counties of Northeastern Ohio and Southeastern Michigan. The counties represent different population densities and demographics, e.g., Cuyahoga (OH) and Wayne (MI) include dense urban areas (Cleveland and Detroit), Summit (OH) and Genesee (MI) represent mid-size urban centers (Akron and Flint), Lake (OH) and Oakland (MI) represent areas near urban centers, while Holmes (OH) and Sanilac (MI) host rural communities.
The data consist of the confirmed daily cases made available by USAfacts \cite{USAfacts}.

The values of the model parameters and of the prior densities, as described in the algorithm in Section~\ref{sec:PF}, are listed in  Table~\ref{tab:params}. 
The innovation covariance matrix $\mC$ is a diagonal matrix of the form
\[
 \mC = {\rm diag}(\sigma_1^2/N^2, \sigma_1^2,\sigma_1^2,\sigma_1^2,\sigma_2^2),
\] 
where $\sigma_1^2$ and  $\sigma_2^2$ are the variances of the uncertainty in the logarithm of the state vector $Z_t$ and in the logarithm of the transmission rate $\beta_t$,respectively. Observe that for the susceptible population, the variance is weighted with the population squared to keep the innovation for this cohort from becoming excessive. We have
\[
    S_{t+1}^j =\widehat S_{t+1}^j {\rm exp}\left(\frac{\sigma_1}{N}w_{t+1}^j\right)\approx \widehat S_{t+1}^j  +\sigma_1\frac{\widehat S_{t+1}^j }{N} w_{t+1}^j,
\]
and without the scaling by $N$, the innovation in a large population could be significantly large. Numerical tests indicate that with a too large innovation, the algorithm may go astray.  
    
\begin{table}
\centerline{
\begin{tabular}{| l|l|l |}
\hline
{\large{\bf Particle filter paremeters}} & & \\
\hline
a priori lower bound transmission rate [1/days] & $\beta_{\rm min}$ & 0.1 \\
a priori upper bound for transmission rate & $\beta_{\rm max}$ & 0.5 \\
a priori upper bound of initial infectious individuals  & $\lambda_{\rm max}$ & 5 \\
number of particles & $N$ & 20\,000 \\
standard deviation of innovation of the state & $\sigma_1$ & 0.01 \\
standard deviation of innovation of the transmission rate &  $\sigma_2$ & 0.1 \\
\hline
\hline
{\large{\bf Model paremeters}} & & \\
\hline
recovery rate & $\gamma$ & 1/21 \\
incubation rate & $\eta$ & 1/7 \\
death rate & $\mu$ & 0.004 \\
\hline
\end{tabular}
}
\caption{\label{tab:params} The parameter values used in the calculations}
\end{table}

\begin{figure}[ht]
\centerline{
\includegraphics[width=18cm]{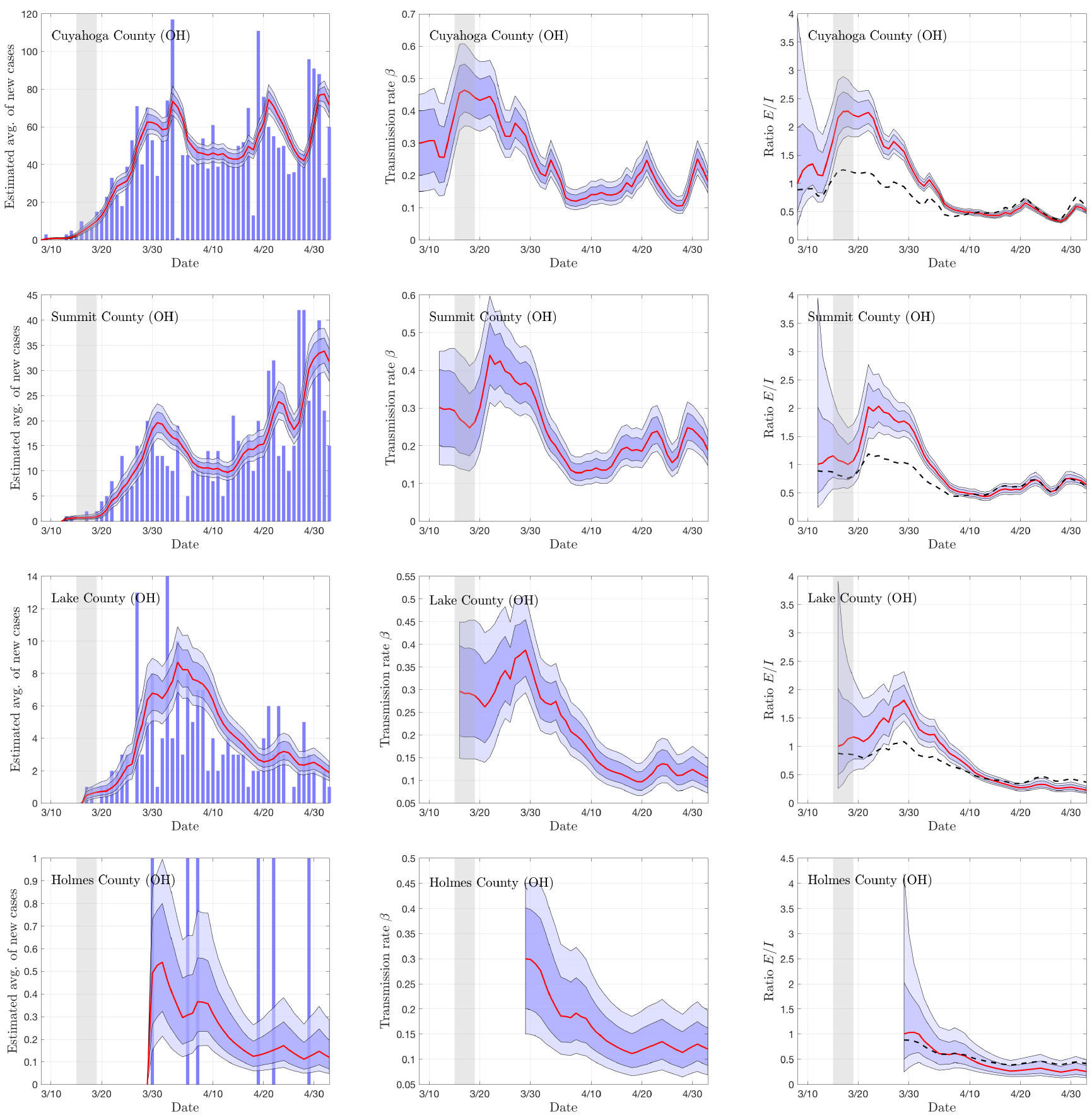}
}
\caption{\label{fig:estimates OH} Particle filter estimates with  of 50\% and 75\% belief  posterior envelopes using the COVID-19 data of four counties in Northeastern Ohio. The top row corresponds to Cuyahoga County (Cleveland area), the second row to Summit County (Akron area), the third row to Lake County (commuter community close to Cleveland), and the bottom row to the rural Holmes County. In the left column, the blue bars indicate the daily count of confirmed cases and the grey bars are the deaths (not used here), while the envelopes indicate the nowcasting of the mean of new cases with model uncertainties. The red curve indicates the median of the sample. In the middle column, the posterior envelopes of 50\% and 75\% of the transmission rate $\beta$ are shown. On the right, we show the posterior envelopes of the ratio of the cohorts $E$ and $I$. The dashed line indicates the equilibrium value $\rho^*$ based on the Riccati equation, computed with the posterior mean of $\beta$, and with the approximation $S/N\approx 1$.}
\end{figure}

\begin{figure}[ht]
\centerline{
\includegraphics[width=18cm]{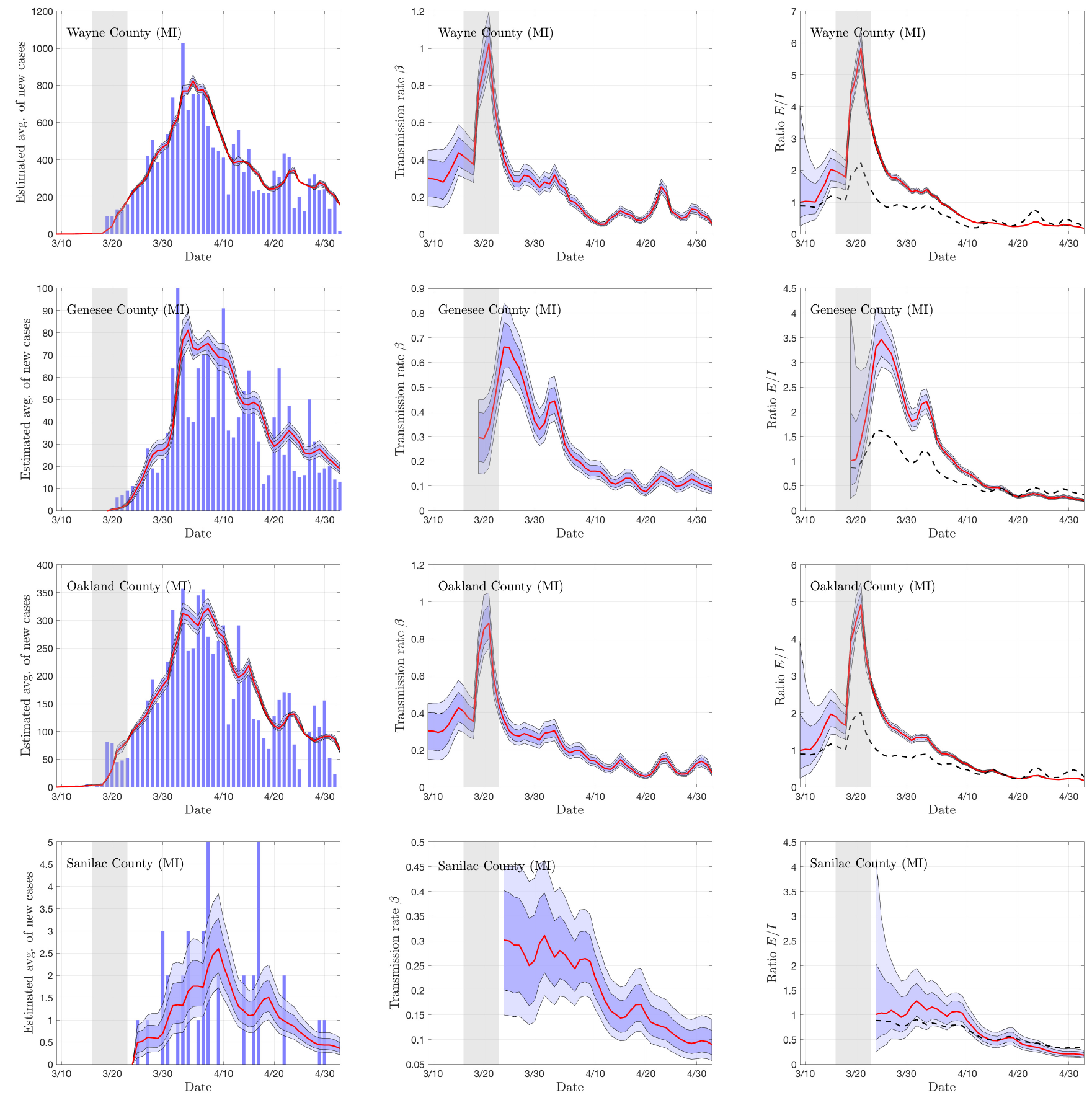}
}
\caption{\label{fig:estimates MI} Particle filter estimates 50\% and 75\% belief posterior envelopes using the COVID-19 data of four counties in Southeastern Michigan. The top row corresponds to Wayne County (Detroit area), the second row to Genesee County (Flint area), the third row to Oakland County (suburban near Detroit), and the bottom row to the rural Sanillac County. For the explanation of the different columns, see the caption of Figure~\ref{fig:estimates OH}.}
\end{figure}

The panels in the left column of Figures~\ref{fig:estimates OH}-\ref{fig:estimates MI} display the raw data and the dynamically estimated expected values of the new infections computed with the particle filter.
In the estimation process, we averaged the data over a moving window of three days to attenuate the effects of, e.g.,  reporting lags, and differences between weekends and weekdays. For each county we show the 50\% and 75\% credibility intervals and the median defined by the particles. More precisely, for each parameter/state vector sample $X_t^j$, $1\leq j\leq N$, we calculate the corresponding fluxes $\varphi_2^j(t) = \eta E_t^j$, and evaluate the interval obtained by discarding the $1/2\times(1-p/100)\times N$ smallest and largest values. The red curve is the median value of the fluxes.

The middle column in Figures~\ref{fig:estimates OH}-\ref{fig:estimates MI} shows the  50\% and 75\% posterior belief envelope of the transmission rate parameter for the respective counties.

Finally, the right column of Figure~\ref{fig:estimates OH}-\ref{fig:estimates MI} shows the 50\% and 75\% posterior envelopes estimated ratio between the cohorts $E$ and $I$. In the same figure, the dashed curve represents the equilibrium value  $\rho^*$ of the ratio based on the Riccati equation, computed from equation (\ref{equilibrium}) with $\beta(t)$ equal to the posterior mean,
\[
 \overline \beta(t) = \sum_{j=1}^N w_t^j \beta_t^j.
\]
Towards the end of the observation period the equilibrium value is in good agreement with the sample-based estimate. 

\begin{figure}[ht]
\centerline{
\includegraphics[height=6cm]{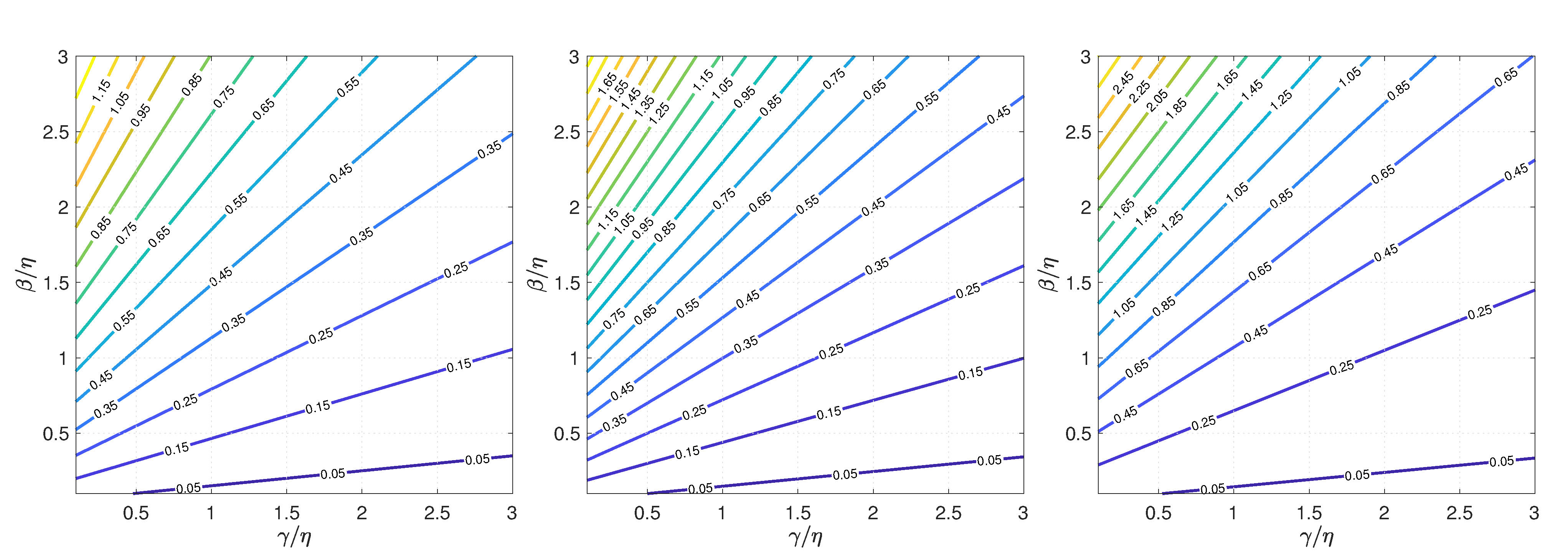}
}
\caption{\label{fig:equilibrium} The equilibrium value $\rho^*$ of the ratio $E/I$ as a function of the dimensionless quantities $t = \gamma/\eta$ and $s = \beta/\eta$ with three different values of the infectivity factor of the cohort $I$: $q=0.1$ (left), $q=0.5$ (center) and $q=1$ (right).}
\end{figure}

The numerical results indicate that the asymptotic value $\rho^*$ is a good proxy of the ratio $E/I$, in particular once the transmission rate $\beta$ has stabilized, therefore providing a quick way to assess the size of the asymptomatic cohort from the size of infected cohort.  To understand better the dependency of the ratio $\rho^*$ on the model parameters, we introduce two dimensionless quantities characterizing the system of differential equations,
\[
 t = \frac \gamma\eta,\quad s =\frac \beta\eta.
\]
Neglecting the effect of the death rate $\mu$, and assuming that the infection is not yet widespread, so that $S/N\approx 1$, we find an approximate formula for $\rho^*$  of (\ref{equilibrium}) in terms of the dimensionless quantities $t$ and $s$,
\begin{equation}\label{equilibrium2}
 \rho^* =   -\frac 12( 1+2t  - sq) + \sqrt{\frac 14 (1+2t - s q)^2
 +s p},
\end{equation}
Figure~\ref{fig:equilibrium} shows the equilibrium value as a function of the dimensionless parameters for three different choices of the parameter: $q=0.1$ (left), $q=0.5$ (center) and $q=1$ (right). In the four Ohio counties, the equilibrium value is consistently near $\rho^*\approx 0.5$, while slightly below it in the Michigan counties. Observe that the value $\rho^*=0.5$ would correspond to effective basic reproduction number
\[ 
R_0 \approx \frac{1/7}{1/21} \rho^* \approx 1.5,
\]
indicating that the disease is in a slow progression phase.

\begin{figure}
\centerline{
\includegraphics[width=18cm]{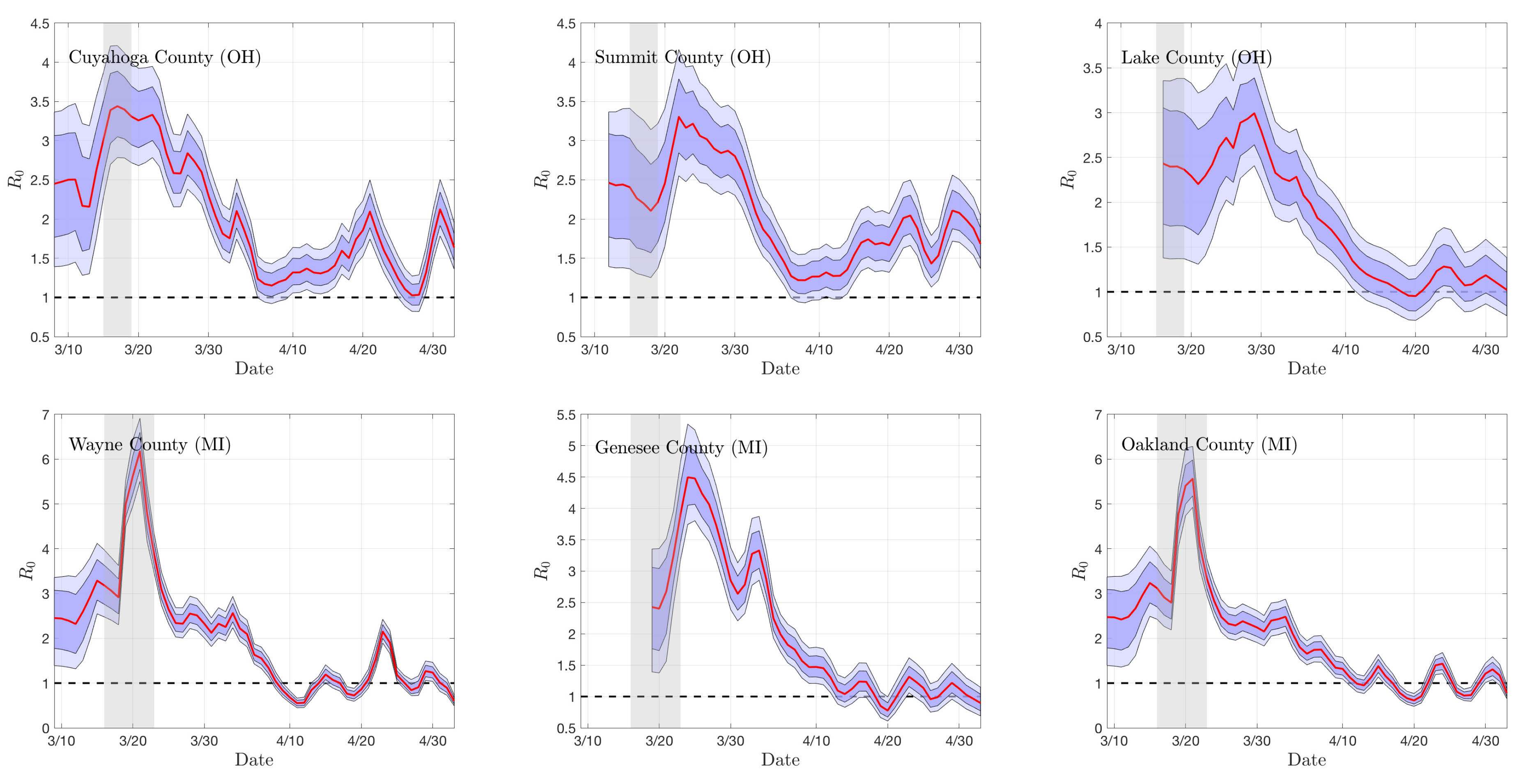}
}
\caption{\label{fig:R0}The estimated $R_0$ posterior envelopes for three counties in Ohio (top row), and four in Michigan (bottom row). The lighter envelope correspond to 75\% posterior belief, and the darker one to 50\% belief.}
\end{figure}

While this definition of $R_0$ in terms of $\rho^*$ implicitly assumes equilibrium, it is possible to compute the time course of $R_0$ for the period when the system has not reached the equilibrium, by using formula (\ref{equilibrium2}) and the estimated transmission rate $\beta$. Figure~\ref{fig:R0} shows the posterior belief envelopes for $R_0$ for six counties, three in Ohio, and three in Michigan, calculated from the posterior sample for $\beta$ at each time. We did not show the results for Holmes County (OH) and Sanilac County (MI), whose $R_0$ was consistently very low. The time courses of $R_0$  show similar patterns in all six counties,
and while in Michigan the peak values were higher than in Ohio, the values towards the end of the observation period are lower, varying around the critical values $R_0=1$.

\begin{figure}[ht]
\centerline{
\includegraphics[width=18cm]{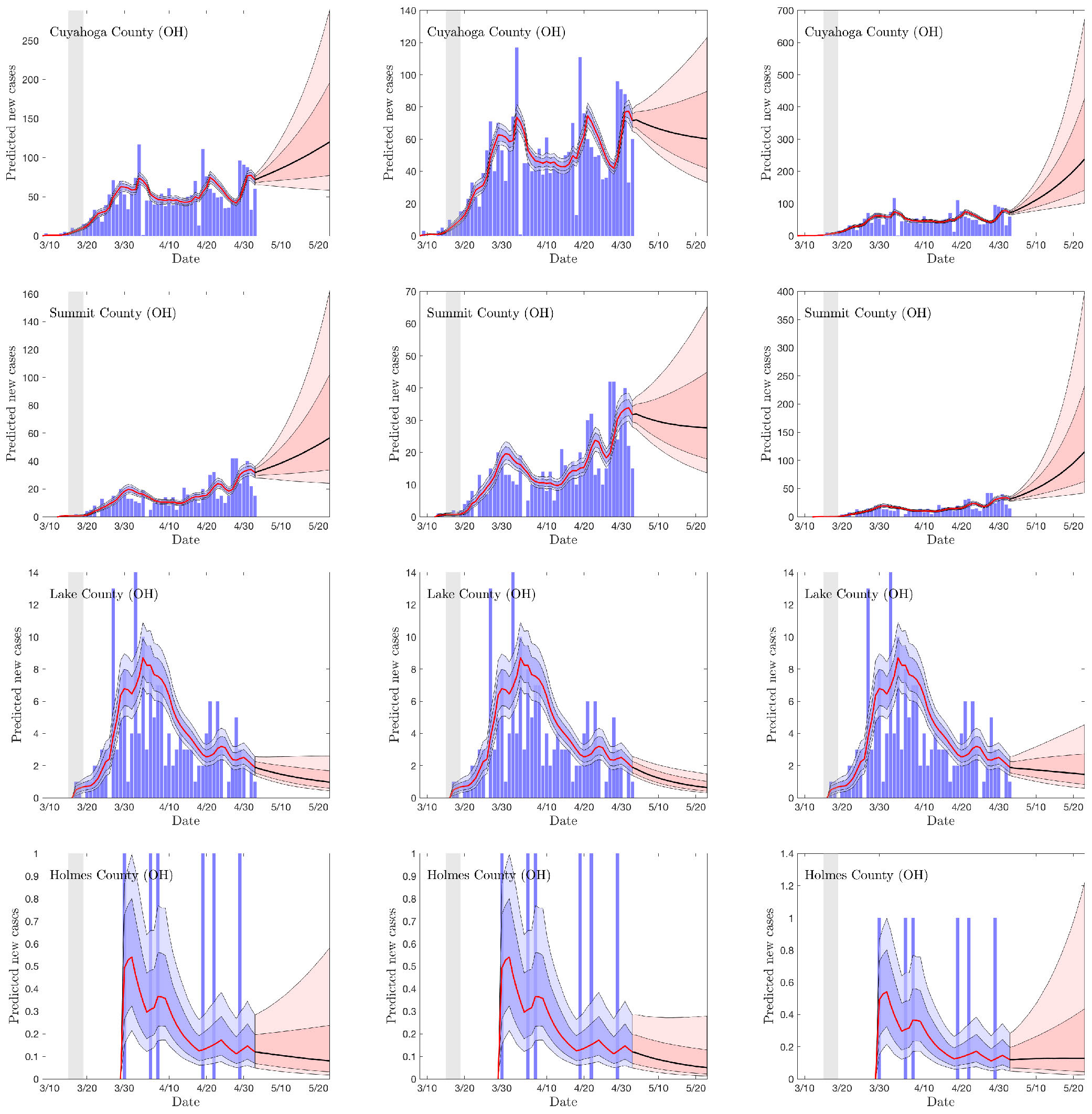}
}
\caption{\label{fig:predictions OH} The predictive envelopes for four counties in Ohio. In the left column, the propagation for each particle is continued by using the last state vector as initial data, and the last estimated $\beta$ value as transmission rate. In the middle column, the transmission rates are slightly lowered by a factor of 0.8, while in the right column, the rates are increased by a factor of 1.2. The predictive envelopes correspond to 50\% and 75\% levels of belief.} 
\end{figure}

\begin{figure}[ht]
\centerline{
\includegraphics[width=18cm]{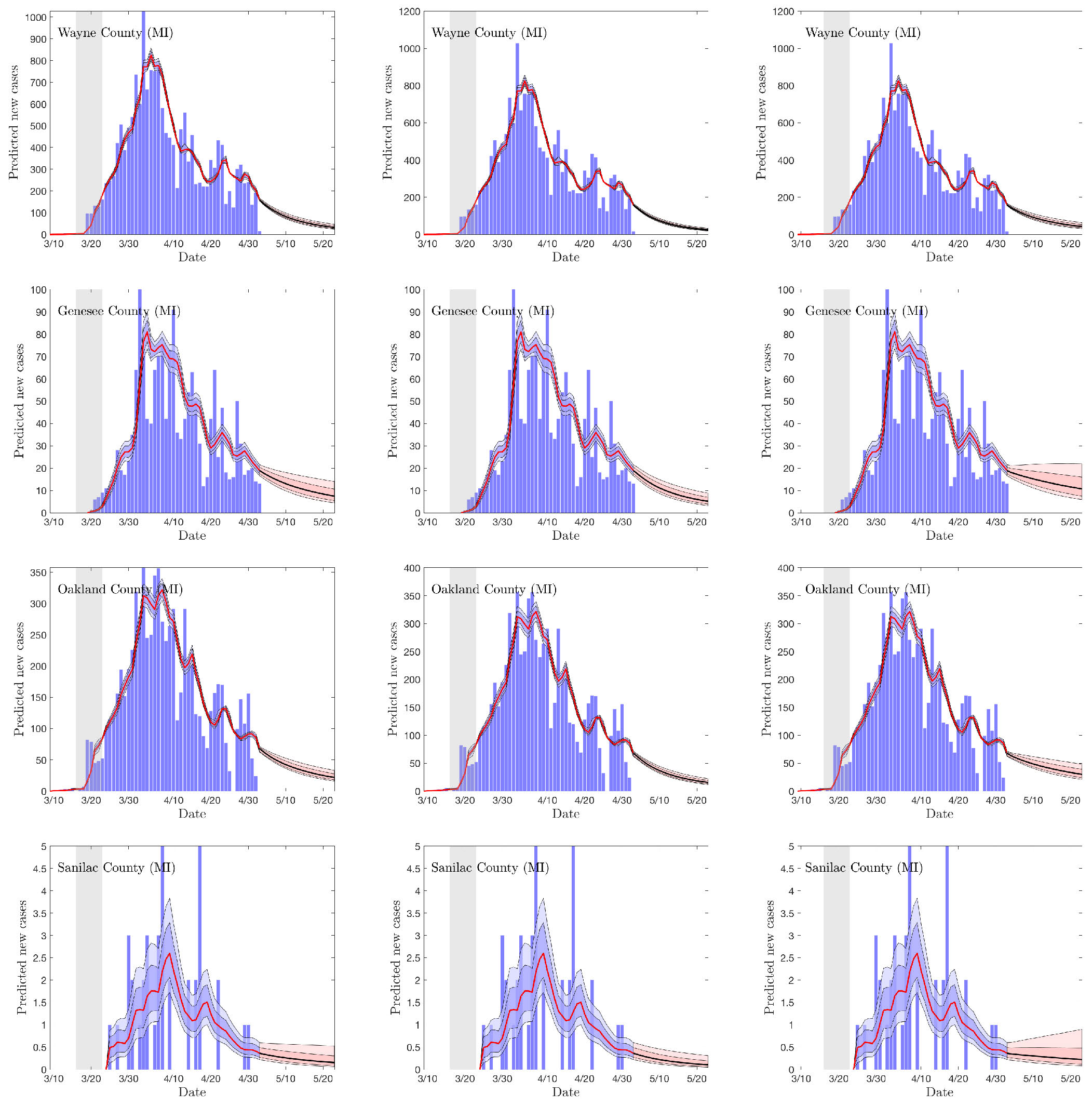}
}
\caption{\label{fig:predictions MI} The predictive envelopes for four selected counties in Michigan. The explanations of the columns are the same as in Figure~\ref{fig:predictions OH}.} 
\end{figure}

The particle solutions to the estimation problem provide a natural tool for predicting new infections. For each particle, the last realization $x_t^j$ provides an initial value for the differential equations and an estimated value for the transmission rate, thus we may use the system of ODEs to propagate the particle data forward in time. From the ensemble of the predictions we can then extract the forecasting belief envelopes, similarly as we did for the nowcasting problem. Figures~\ref{fig:predictions OH}-\ref{fig:predictions MI} show the predictions for the following 20 days for the four sample counties under the assumption that the mitigation conditions remain unchanged. In addition, we have run two alternative scenarios: In the first one, the last estimate of the transmission rate for each particle is multiplied by a factor $\kappa = 0.8$, simulating a tightening of the containment and control measures, while in the second scenario, the transmission rate is increased by multiplicative factor $\kappa = 1.2$, simulating a relaxation of the control measures. We observe that the predictions illustrate well the what the lower $R_0$ in Michigan compared to Ohio means in terms of expected number of new infections.

To further test the predictive skills of the model, we leave out from the current data set $\{b_1,b_2,\ldots,b_T\}$ the last $t_{\rm pred}$ data points, considering only the data up to $\widetilde T = T - t_{\rm pred}$, and compute a sample of predicted new cases, defining the predicted particle sample
\[
 b_{{\rm pred},t}^j \sim{\rm Poisson}(\eta E_ {{\rm pred},t}^j),\quad  t = \widetilde T +1, \ldots, \widetilde T + t_{\rm pred} = T,  \quad 1\leq j\leq N,
\]
that is, we propagate the particles computed at $\widetilde T$ forward, and generate artificial observations. A comparison of the resulting predictive envelopes and the actual data is indicative of the predictive skill of the model: large deviations indicate that the model may not have taken some important factors into account. 

\begin{figure}
\centerline{
\includegraphics[width=18cm]{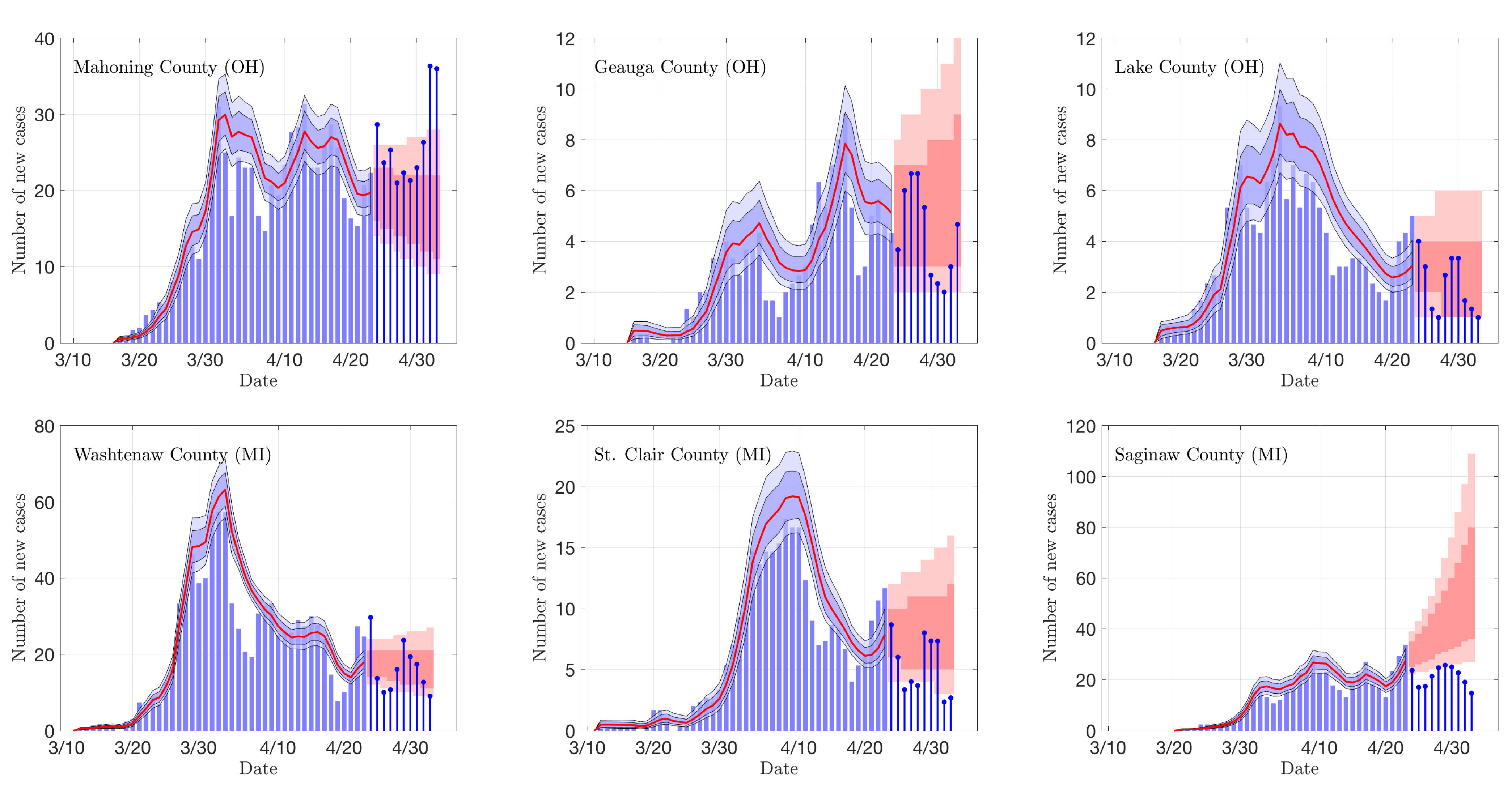}
}
\caption{\label{fig:predictive}Ten days prediction of new daily infections in six different counties, three in Ohio and three in Michigan. The predictions are based on the data excluding the last ten observations. The two different shades correspond to 50\% and 75\% uncertainty of predicting new observations. The actual data are shown as stem plots.}
\end{figure}

Figure~\ref{fig:predictive} shows the $10$ day predictions for three counties in Ohio and three counties in Michigan,  together with the actual data. The plots indicate that, in general, the model captures the trend of the data relatively well, with the exception of the Saginaw County (MI), where the predicted trend seems higher than the actual observation.
% A more comprehensive set of examples can be found in the supplementary material.

\begin{figure}
\centerline{
\includegraphics[width=18cm]{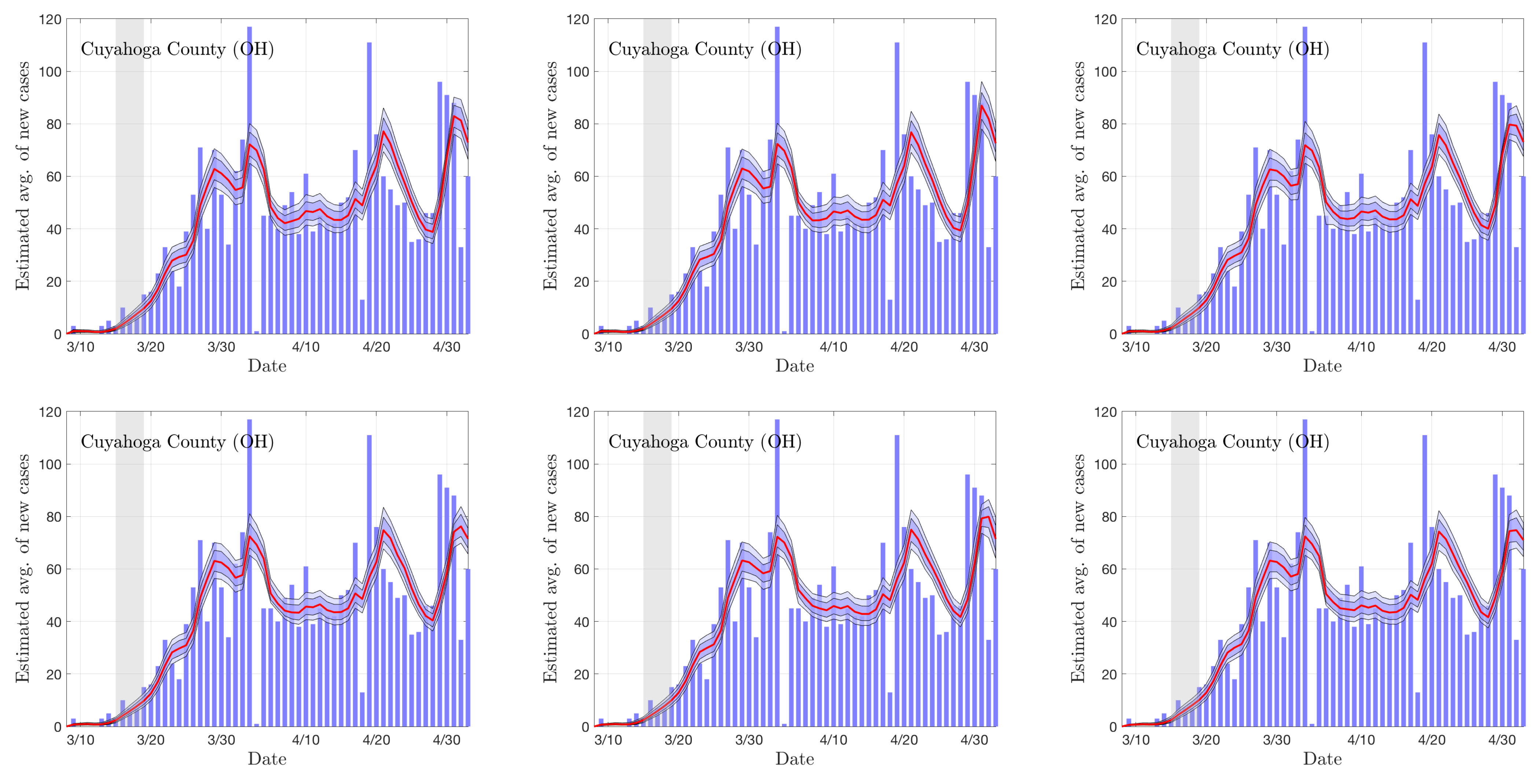}
}
\caption{\label{fig:vary params1} The effect of parameters $\eta = 1/T_{\rm inc}$ and $\gamma = 1/T_{\rm rec}$  on the estimated expected value of new cases. The parameter combinations $T_{\rm inc}/T_{\rm rec}$ in these figures are chosen as follows: First row:
4/7, 4/10, 4/14. Second row: 7/10, 7/14, 7/21.}
\end{figure}

\begin{figure}
\centerline{
\includegraphics[width=18cm]{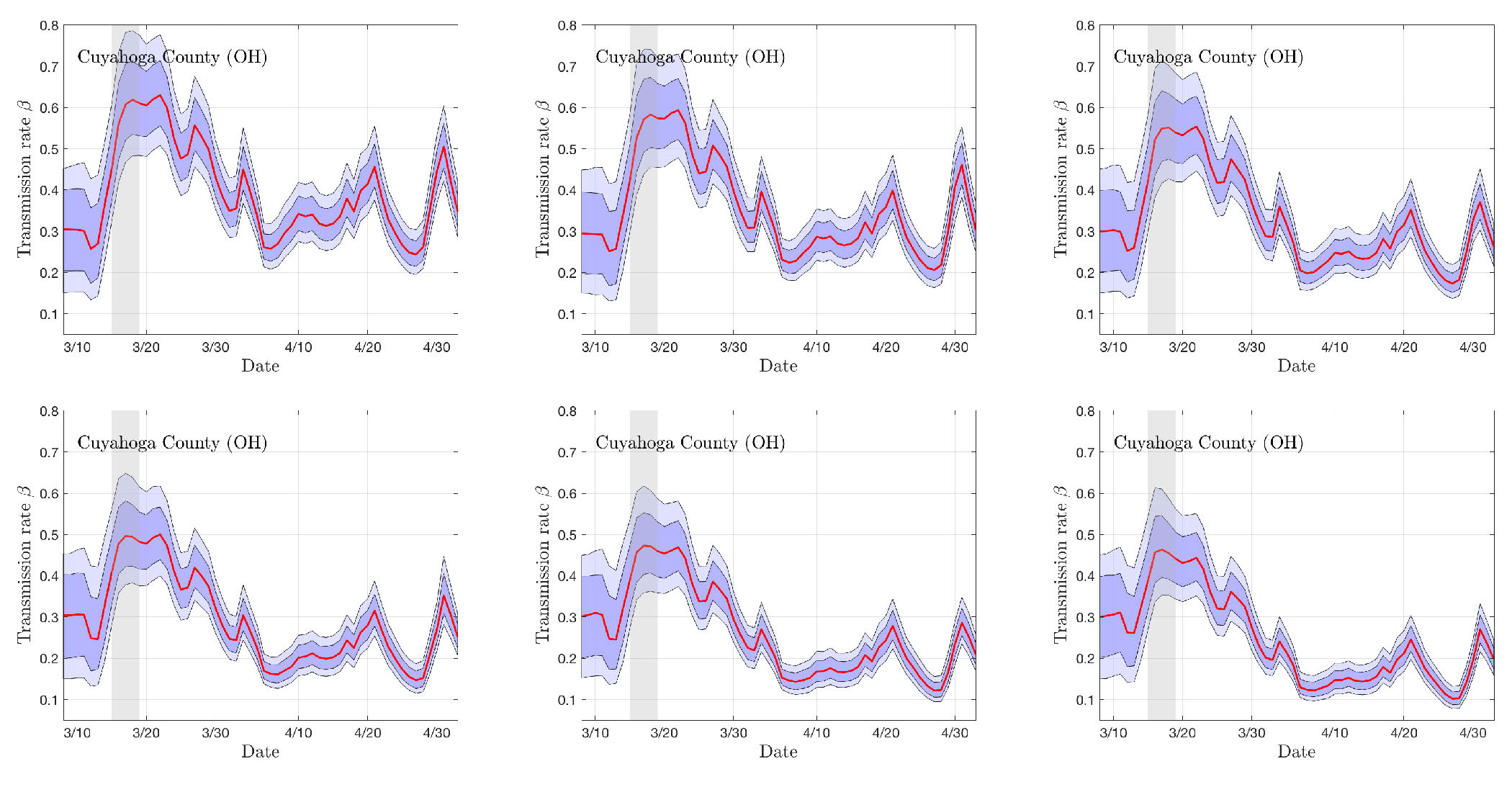}
}
\caption{\label{fig:vary params2} The effect of parameters $\eta = 1/T_{\rm inc}$ and $\gamma = 1/T_{\rm rec}$  on the estimated value of the transmission rate $\beta$. The parameter combinations $T_{\rm inc}/T_{\rm rec}$ in these figures are chosen as follows: First row:
4/7, 4/10, 4/14. Second row: 7/10, 7/14, 7/21.}
\end{figure}

\begin{figure}
\centerline{
\includegraphics[width=18cm]{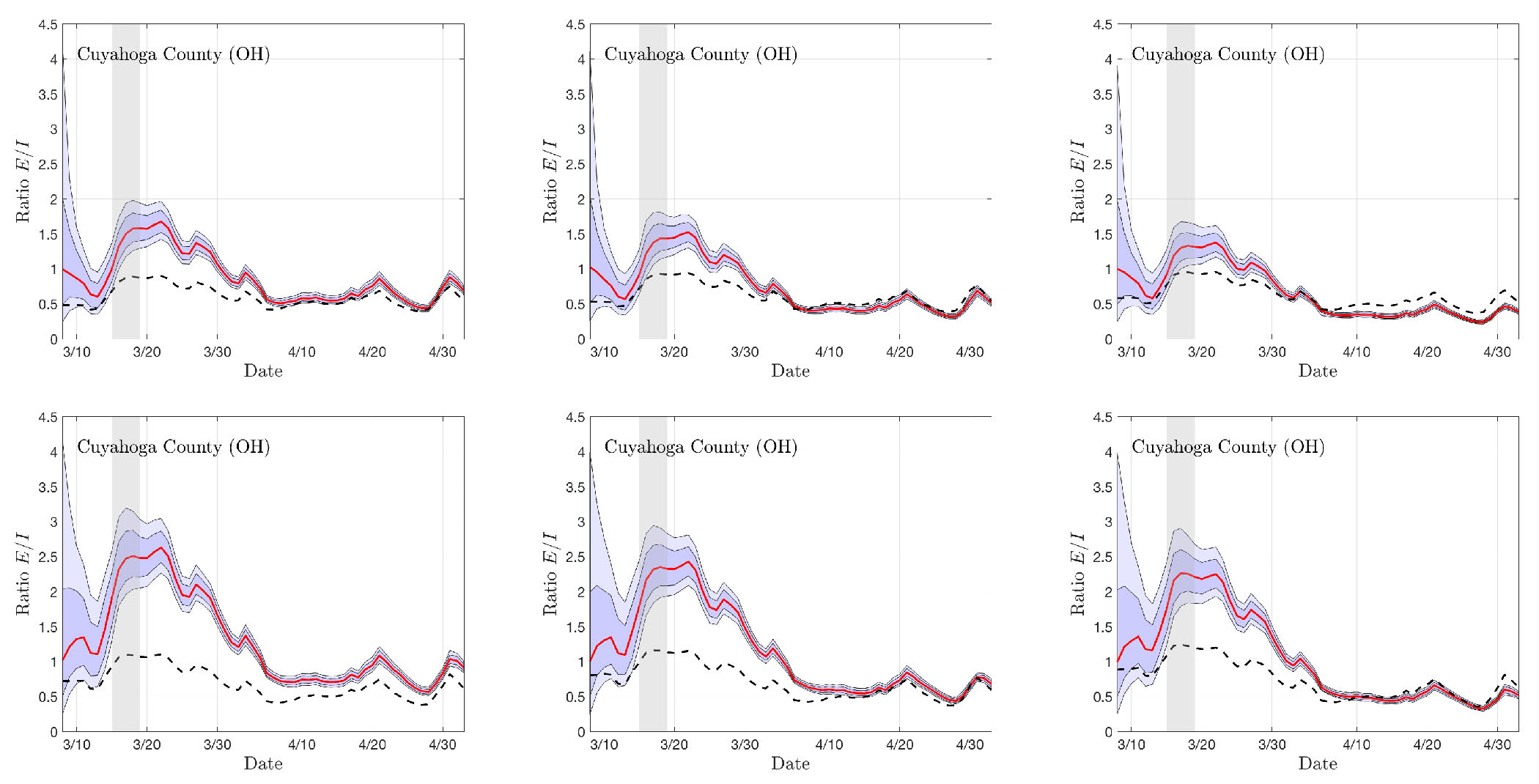}
}
\caption{\label{fig:vary params3} The effect of parameters $\eta = 1/T_{\rm inc}$ and $\gamma = 1/T_{\rm rec}$  on the estimated value of the ratio $E(A)/I$. The equilibrium value $\rho^*$ is plotted as a dashed curve. The parameter combinations $T_{\rm inc}/T_{\rm rec}$ in these figures are chosen as follows: First row:
4/7, 4/10, 4/14. Second row: 7/10, 7/14, 7/21.}
\end{figure}

\begin{figure}
\centerline{
\includegraphics[width=18cm]{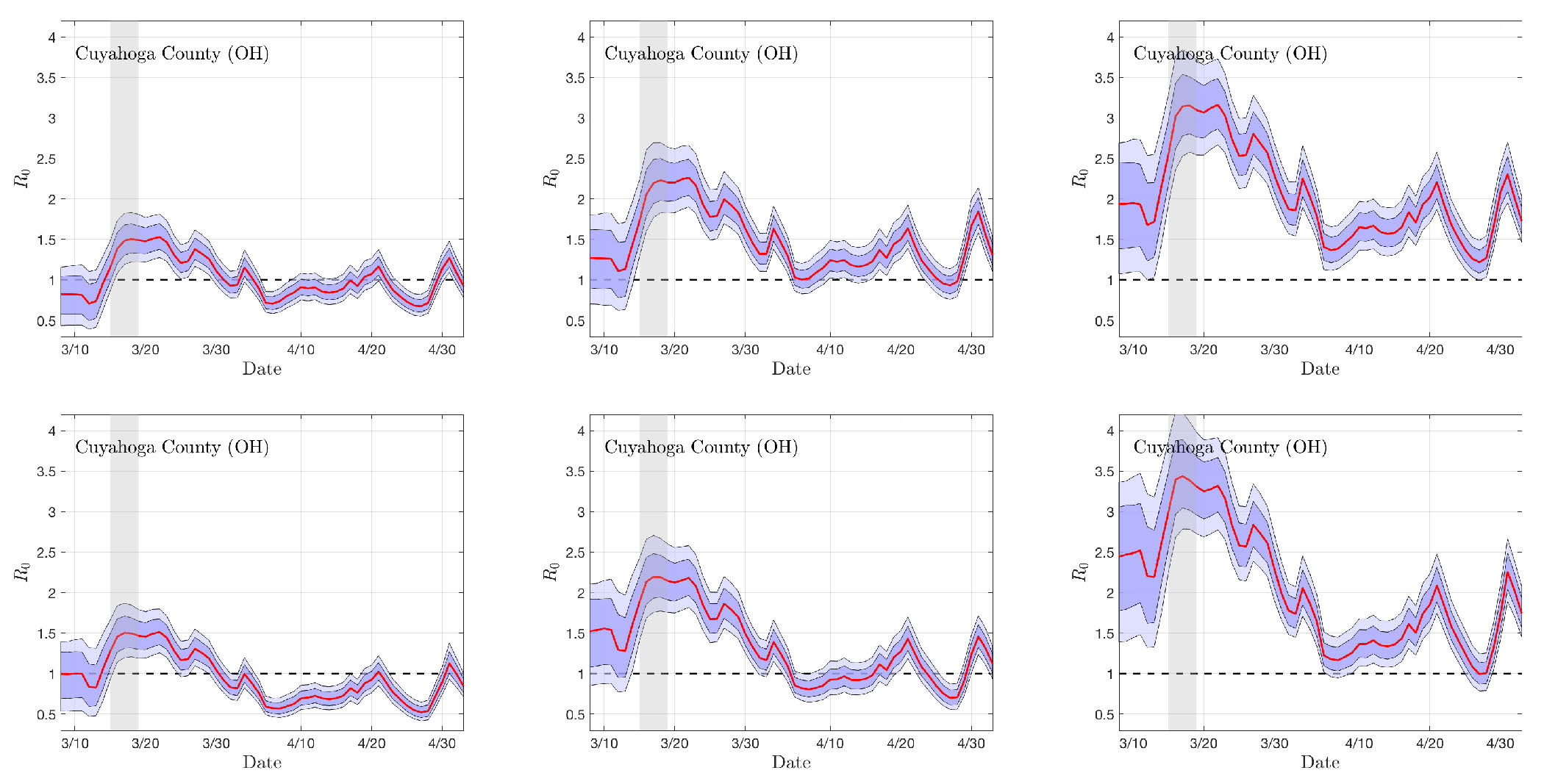}
}
\caption{\label{fig:vary params4} The effect of parameters $\eta = 1/T_{\rm inc}$ and $\gamma = 1/T_{\rm rec}$  on the estimated reproduction number $R_0$. The parameter combinations $T_{\rm inc}/T_{\rm rec}$ in these figures are chosen as follows: First row:
4/7, 4/10, 4/14. Second row: 7/10, 7/14, 7/21.}
\end{figure}

Finally, we consider the sensitivity of the results to the choice of some of the model parameters that assumed to be known, and in particular, the time constants $T_{\rm inc} = 1/\eta$  (incubation) and $T_{\rm rec} = 1/\gamma$ (recovery). To test the sensitivity, we select one of the counties, Cuyahoga (OH) and run the program with different parameter combinations. Figure~\ref{fig:vary params1}--\ref{fig:vary params4} show a selection of results. The results show that both the estimated transmission rate and the ratio $\rho = E/I$ are rather robust for variations of the parameter, the former decreasing slightly with increasing $T_{\rm rec}$. Interestingly, towards the end of the observation interval, the ratio $\rho$ is in all cases close to the equilibrium value $\rho^*\approx 0.5$. Therefore, the value of $R_0$ is approximately
\[
 R_0 = \frac{\eta}{\gamma+\mu}\rho^* \approx \frac{\eta}{\gamma}\rho^* \approx 0.5\times\frac{T_{\rm rec}}{T_{\rm inc}}.
\] 
However, this approximation gives an idea of the spreading of the transmission only if the ratio $E/I$ is near the equilibrium. The plots show that in particular at the beginning of the observation period the ratio is far from the equilibrium, and $R_0$ would predict a too slow growth rate for the transmission dynamics.

\section{Discussion}

The proposed dynamical Bayesian filtering algorithm based on particle filters is shown to be able to produce a robust and consistent estimate of the state vectors consisting of the sizes of the four cohorts of a new SE(A)IR model for COVID-19 spread, as well as of the transmission rate,  a key parameter that is not  expected to be a constant. In epidemiology models, the transmission rate $\beta$ is usually defined as
\[
 \beta = \mbox{(transmission probability per contact )$\times$(number of contacts per day),}
\]
where the first factor is related to the characteristics of the pathogen, while the second factor is related to the behavior of the individuals and can change in connection with
hygiene, social distancing and other mitigation measures. In the Ohio and Michigan counties, we observe a rapid growth of $\beta$ at the beginning of the outbreak, followed by a clear and consistent drop. The initial increase may be due, to some extent, to the scarcity of the data in the beginning of the epidemics, and reflect the learning effort of the  algorithm. The drop that follows, however, correlates well with the stay-at-home order and other social distancing measures introduced early on in both states, from March 16 to March 23 in Michigan, and from March 15 to March 19 in Ohio. In both states and across all counties considered here, the transmission rate is fairly stable, except for  Wayne County, Michigan, where $\beta$ is slightly higher early in the epidemic. This may be the effect of a wider use of public transportation, socio-economic conditions and demographic factors. If the reason for the drop is the adoption of mitigation measures, it is natural to wonder why its onset is not the same in all counties. A possible explanation of the delay may lie in the fact that the infection arrived in different communities through the mobility of infectious individuals, and what happens earlier in large, densely populated areas affects the surrounding communities with a delay. The effects of the network structure on the timing and intensity of the COVID-19 spread will be addressed separately.

This SE(A)IR model we present directly addresses the role of asymptomatic individuals shedding virus, many of whom recover before developing symptoms, in transmission dynamics.  We demonstrate a method for dynamically estimating the ratio of  asymptomatic/presymptomatic/oligosymptomatic individuals in the $E$ compartment to symptomatic individuals in the $I$ compartment.  This ratio, whose equilibrium value consistently settles to a value less than one across all counties examined and with a wide range of model parameters, is considerably lower than the 50-85 ratio speculated in \cite{bendavid2020covid} based on a serosurvey conducted in California on April 3-4, 2020.  
This disparity may stem partly from the growing availability of testing throughout April as well as the cross-reactivity in the ELISA tests used as part of the serosurvey. Understanding better this discrepancy will be a topic of future studies. However, we emphasize that while our methodology predicts that the asymptomatic cohort is typically smaller than the symptomatic one, according to the model, it is still responsible for the majority of transmissions.
Furthermore, as already pointed out, the current SE(A)IR model does not address some of the features of COVID-19 transmission, including the latent period of the asymptomatic cohort. While designing a model that properly addresses the delay of the infectious phase of the asymptomatic patients is straightforward, 
retaining a structure that allows us to inform about the size of asymptomatic cohort based on data on symptomatic patients alone remains a challenge. Elaborating on that point will be a future direction of this research.

The model-based predictions indicate that the current value of $\beta$ alone is not enough to project how the infection process will proceed. In fact, while the value $\beta$ at the end of the considered time interval is nearly the same for all counties, the predictions on the new infections differ significantly. The main reason for this discrepancy is the different size of the cohorts. The numbers infected tend to increase strongly even if $\beta$ is relatively small if the pipeline of exposed/asymptomatic individuals is crowded. 
Because an unobservable concentration of individuals in the exposed/asymptomatic category may represent an uptick in future cases, only several days (preferably two weeks) of consecutively falling case numbers likely represents systematic decline in population-level disease activity.  The uncertainty associated with undulating case counts is reflected in the wide predictive envelopes seen in projections for several of the Ohio counties. It is worth noting 10 day projections shown in Figure~\ref{fig:predictive} tend to overestimate observed incidence in the last few days.  We suspect this is attributable in part to reporting delays.

In this paper we introduce a new way of estimating the reproduction number $R_0$ for the proposed  model, based on an equilibrium condition of a related non-linear Riccati type equation. The equilibrium condition, in turn, was shown to correspond well to the estimated ratio of the symptomatic and non-symptomatic cases after the transmission rate had stabilized, presumably due to the social distancing measures. Interestingly, the equilibrium value $\rho^*$ and the corresponding $R_0$ are very consistent for a number of counties in two states, Ohio and Michigan, that adopted similar distancing measures around the same time, regardless of the fact that Michigan had a significantly higher number of infection than Ohio at the time of the adoption. 
These observations give an indication that the quantities reflect well the effectiveness of the mitigation measures, providing a useful and quick indicator for assessing such measures.

\section*{Acknowledgements}

The work of DC was partly supported by NSF grants DMS-1522334 and DMS-1951446, and the work of ES was partly supported by NSF grant DMS-1714617.

\bibliographystyle{unsrt}
\bibliography{References_PE}

\end{document}